%% file: poster.tex
\documentclass[letterpaper,twocolumn,10pt]{article}
\usepackage{poster}
\usepackage{import}
\usepackage{tikz}
\usepackage{amsmath}

\usepackage{filecontents}

\usepackage{todo}

\begin{document}

\date{}

\title{\Large \bf Routing over QUIC:\\Bringing transport innovations to routing protocols}

\author{
{\rm Thomas Wirtgen\thanks{Student author, supported by an FRS-FRIA grant}}\\
UCLouvain
\and
{\rm Nicolas Rybowski\thanks{Student author}}\\
UCLouvain
\and
{\rm Cristel Pelsser}\\
UCLouvain
\and
{\rm Olivier Bonaventure}\\
UCLouvain\\
\hspace{-12cm}
\vspace{-0.5cm}
{\rm firstname.lastname@uclouvain.be}
} %

\maketitle

\begin{abstract}
By combining the security features of TLS with the reliability of TCP, QUIC opens new possibilities for many applications. We demonstrate the benefits that QUIC brings for routing protocols.
Current Internet routing protocols use insecure transport protocols. BGP uses TCP possibly with authentication. OSPF
uses its own transport protocol above plain IP. We design and implement a library that allows to replace the transport protocols
used by BGP and OSPF with QUIC. We apply this library to the BIRD routing daemon and report preliminary results.
\end{abstract}

\vspace{-0.4cm}

\section{Introduction}

\vspace{-0.3cm}

Internet routing protocols exchange different types of messages. For historical reasons, the
intradomain routing protocols (IS-IS and OSPF) have defined their own protocol to provide a
reliable delivery over point-to-point links or LANs. Over the years, several limitations have
been identified in these specialized protocols. First, they did not include any security features and 
relied on password-based authentication. Newer versions support hash-based authentication techniques
or support IPSec tunnels between routers. Second, their performance is not equivalent to the performance
of modern reliable transport protocols. There are ongoing discussions within the IETF to better tune these
specialized transport protocols~\cite{ietf-lsr-isis-fast-flooding-02}.

When BGP was designed, its inventors identified the need for a reliable transport protocol and decided to use TCP, although they noted
that TCP could be replaced by an equivalent protocol. To our knowledge, such a replacement has not yet been 
fully evaluated. BGP over TCP has various limitations. First, it does not provide security features that are
required to protect BGP sessions from attacks. Operators have deployed various techniques, ranging from using a TTL of 255
on single hop eBGP sessions to hash-based authentication with TCP-MD5 and TCP-AO.

In this poster, we experiment with the BIRD BGP and OSPF implementations to evaluate whether it is possible to replace
the existing transport protocols with QUIC.
We first propose a socket API which can expose QUIC to a routing daemon.
We report preliminary evaluation demonstrating that BGP over QUIC is not significantly slower than BGP over TCP while being obviously more secure. 
We also report our experience with using OSPF over QUIC.

\vspace{-0.4cm}

\section{Socket API}\label{section:api}

\vspace{-0.3cm}

Current Internet routing protocols have been strongly coupled with their underlying transport protocols. We envision that future Internet routing protocols would be less coupled with the transport layer and could negotiate the utilization of different transport protocols. 
To demonstrate the feasibility of this approach, we designed
a socket API abstracting the main features of recent transport
protocols. Since most QUIC implementations do not propose a common interface,
we created a "socket-like" common layer that all QUIC implementations can
follow. %
Our API consists of 4737 Lines of Code (LoC) and supports multiple QUIC implementations,
the main ones being \texttt{picoquic} and \texttt{MsQuic}.
We have extended the BIRD routing stack to incorporate this API.
The total modifications to BIRD to support our QUIC socket API consist
of only 759 LoCs.
These few modifications come from the fact that BIRD also uses a
socket API to communicate with transport implementations in kernel space and thus
many BIRD subroutines could be reused or modified to add our QUIC socket API.

\vspace{-0.4cm}

\section{BGP over QUIC}\label{section:boq}

\vspace{-0.3cm}

\begin{figure}[t!]
    \centering
    \def\svgwidth{6.5cm}
    \footnotesize
    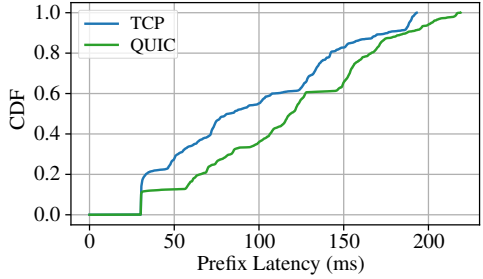
    \caption{Latency to propagate prefixes in the testing BGP network.}
    \label{fig:bgp_delay}
\end{figure}

BGP uses TCP as its reliable transport protocol.
However, RFC1105 stated that ``\textit{any reliable transport protocol may be used}''.
Since then, no BGP extension was proposed to use another transport protocol than TCP.
The first part of this work is to replace TCP with QUIC to take advantage of its
security benefits, but also the features it offers. 
Figure~\ref{fig:bgp_delay} shows that the introduction of QUIC has a relatively small impact on performance compared to TCP.

To evaluate the performance impact of the
new transport stack, we created a topology of 4 routers: \texttt{R1}, \texttt{R2} and \texttt{R3} connected in triangle and \texttt{GoBGP} connected to \texttt{R1}.
We use the GoBGP implementation to establish a BGP (over TCP) session with \texttt{R1} and inject a complete
routing table from a RIPE RIS snapshot of \texttt{rrc00} dated from 23 November 2022 at 8 am. 
In total, 970k IPv4 routes and 171k IPv6 routes are injected on \texttt{R1}.
\texttt{R1} monitors all BGP update messages that are received from \texttt{R2} and \texttt{R3}.
Routers \texttt{R1}, \texttt{R2}, and \texttt{R3} are configured to establish a QUIC or TCP session
depending on the experiment. We measure the time it takes for each prefix
to be fully propagated, i.e., from router \texttt{R1} to \texttt{R2}, \texttt{R3}
and finally returned back to \texttt{R1}. Figure~\ref{fig:bgp_delay} shows that using
QUIC does
not significantly delay the propagation of BGP routes compared to TCP, despite
the obvious security benefits of QUIC.

\vspace{-0.5cm}

\section{OSPF over QUIC}\label{section:Ooq}

\vspace{-0.3cm}

The OSPF specification defines a custom transport protocol running directly above IP.
It is based on various timers triggering the main route distribution events, i.e., $(i)$ the creation of a session with the peer(s), $(ii)$ the initial exchange of the link state database (LSDB) and $(iii)$ the periodical exchange of information to keep the distributed LSDBs synchronized. We experimentally measured a convergence time of around 7 seconds upon injection of a new prefix in a full mesh of 6 nodes with plain OSPF.

Our first approach is to modify the transport layer. Instead of exchanging
messages directly over IP, they are sent using a QUIC stream.
This is achieved by slightly changing the OSPF Hello handshake.
When an interface reaches the \texttt{2-Way} state, a QUIC connection is created with the peer.
Once the QUIC session is established, messages from the unmodified OSPF protocol are sent to the peer through it.
The convergence time, evaluated on the 6 nodes testbed, is stable around 12 seconds.
The performance penalty is mainly explained by the fact that we run unmodified OSPF over a stream-based transport protocol. For example, OSPF ACKs are also ACKed by the transport layer, hence consuming useless RTTs.
Fragmentation is still performed by OSPF but should be delegated to QUIC.
By entirely removing the OSPF transport layer and exploiting the advanced features of QUIC, we expect to improve the IGP performances. 

\vspace{-0.4cm}

\section{Discussion and Future work}\label{section:conclusion}

\vspace{-0.3cm}

For both BGP and OSPF, introducing QUIC as the transport layer showed some performance impacts.
Those are explained by multiple reasons.
First, we use the \texttt{picoquic} implementation which is not designed for pure performance but is mainly intended to let developers test new features from the QUIC specifications.
Second, unlike TCP or IP, QUIC is implemented in user space, which involves
more context switching. Third, QUIC establishes an encrypted communication channel for security purposes and therefore performs more operations than a simple TCP connection or IP packet exchange.

Nevertheless, there are security advantages to use QUIC. First, the
data exchanged by the routers is encrypted without the need to configure IPSec tunnels
which can be cumbersome to maintain. Compared to MD5 or TCP-AO where routers exchange a
shared symmetric key, each router now has an X.509 certificate.
Some attacks targeting BGP such as TCP reset or route injection become more difficult because data is exchanged over an encrypted communication channel.

BGP over QUIC is currently being discussed within the IETF~\cite{retana-idr-bgp-quic-00}.
However, our approach is different as we consider that BGP should not
be aware of the internals of the underlying transport protocol.
IGP over TCP and QUIC has also been discussed at the IETF, but no consensus has been reached~\cite{hsmit-lsr-isis-flooding-over-tcp-00}.
To the best of our knowledge, this poster is the first attempt to propose a prototype of Internet routing protocols running on top of a modern transport protocol such as QUIC.

In the next steps, we will improve the integration of the routing protocols with QUIC.
We will also leverage its advanced features like connection migration or multipath capabilities to explore new use-cases.

\vspace{-0.4cm}

\section*{Availability}
\vspace{-0.3cm}

To enable other researchers to experiment with Routing over QUIC and improve it,
we plan to release the source code of our modified version of BIRD with upcoming publications.

\vspace{-0.4cm}

\small
\bibliographystyle{abbrvurl}
\bibliography{\jobname}

\end{document}

%% file: figures/delays_paper.pdf_tex
\begingroup%
  \makeatletter%
  \providecommand\color[2][]{%
    \errmessage{(Inkscape) Color is used for the text in Inkscape, but the package 'color.sty' is not loaded}%
    \renewcommand\color[2][]{}%
  }%
  \providecommand\transparent[1]{%
    \errmessage{(Inkscape) Transparency is used (non-zero) for the text in Inkscape, but the package 'transparent.sty' is not loaded}%
    \renewcommand\transparent[1]{}%
  }%
  \providecommand\rotatebox[2]{#2}%
  \newcommand*\fsize{\dimexpr\f@size pt\relax}%
  \newcommand*\lineheight[1]{\fontsize{\fsize}{#1\fsize}\selectfont}%
  \ifx\svgwidth\undefined%
    \setlength{\unitlength}{303.12000275bp}%
    \ifx\svgscale\undefined%
      \relax%
    \else%
      \setlength{\unitlength}{\unitlength * \real{\svgscale}}%
    \fi%
  \else%
    \setlength{\unitlength}{\svgwidth}%
  \fi%
  \global\let\svgwidth\undefined%
  \global\let\svgscale\undefined%
  \makeatother%
  \begin{picture}(1,0.57719714)%
    \lineheight{1}%
    \setlength\tabcolsep{0pt}%
    \put(0,0){\includegraphics[width=\unitlength,page=1]{delays_paper.pdf}}%
    \put(0.15130332,0.06042291){\color[rgb]{0,0,0}\makebox(0,0)[lt]{\lineheight{1.25}\smash{\begin{tabular}[t]{l}0\end{tabular}}}}%
    \put(0,0){\includegraphics[width=\unitlength,page=2]{delays_paper.pdf}}%
    \put(0.31410881,0.06042291){\color[rgb]{0,0,0}\makebox(0,0)[lt]{\lineheight{1.25}\smash{\begin{tabular}[t]{l}50\end{tabular}}}}%
    \put(0,0){\includegraphics[width=\unitlength,page=3]{delays_paper.pdf}}%
    \put(0.47691431,0.06042291){\color[rgb]{0,0,0}\makebox(0,0)[lt]{\lineheight{1.25}\smash{\begin{tabular}[t]{l}100\end{tabular}}}}%
    \put(0,0){\includegraphics[width=\unitlength,page=4]{delays_paper.pdf}}%
    \put(0.65020966,0.06042291){\color[rgb]{0,0,0}\makebox(0,0)[lt]{\lineheight{1.25}\smash{\begin{tabular}[t]{l}150\end{tabular}}}}%
    \put(0,0){\includegraphics[width=\unitlength,page=5]{delays_paper.pdf}}%
    \put(0.82350502,0.06042291){\color[rgb]{0,0,0}\makebox(0,0)[lt]{\lineheight{1.25}\smash{\begin{tabular}[t]{l}200\end{tabular}}}}%
    \put(0.38192177,0.01531907){\color[rgb]{0,0,0}\makebox(0,0)[lt]{\lineheight{1.25}\smash{\begin{tabular}[t]{l}Prefix Latency (ms)\end{tabular}}}}%
    \put(0,0){\includegraphics[width=\unitlength,page=6]{delays_paper.pdf}}%
    \put(0.04824127,0.11664222){\color[rgb]{0,0,0}\makebox(0,0)[lt]{\lineheight{1.25}\smash{\begin{tabular}[t]{l}0.0\end{tabular}}}}%
    \put(0,0){\includegraphics[width=\unitlength,page=7]{delays_paper.pdf}}%
    \put(0.04824127,0.19936412){\color[rgb]{0,0,0}\makebox(0,0)[lt]{\lineheight{1.25}\smash{\begin{tabular}[t]{l}0.2\end{tabular}}}}%
    \put(0,0){\includegraphics[width=\unitlength,page=8]{delays_paper.pdf}}%
    \put(0.04824127,0.28208601){\color[rgb]{0,0,0}\makebox(0,0)[lt]{\lineheight{1.25}\smash{\begin{tabular}[t]{l}0.4\end{tabular}}}}%
    \put(0,0){\includegraphics[width=\unitlength,page=9]{delays_paper.pdf}}%
    \put(0.04824127,0.36480792){\color[rgb]{0,0,0}\makebox(0,0)[lt]{\lineheight{1.25}\smash{\begin{tabular}[t]{l}0.6\end{tabular}}}}%
    \put(0,0){\includegraphics[width=\unitlength,page=10]{delays_paper.pdf}}%
    \put(0.04824127,0.4475298){\color[rgb]{0,0,0}\makebox(0,0)[lt]{\lineheight{1.25}\smash{\begin{tabular}[t]{l}0.8\end{tabular}}}}%
    \put(0,0){\includegraphics[width=\unitlength,page=11]{delays_paper.pdf}}%
    \put(0.04824127,0.53025169){\color[rgb]{0,0,0}\makebox(0,0)[lt]{\lineheight{1.25}\smash{\begin{tabular}[t]{l}1.0\end{tabular}}}}%
    \put(0.02818939,0.30229929){\color[rgb]{0,0,0}\rotatebox{90}{\makebox(0,0)[lt]{\lineheight{1.25}\smash{\begin{tabular}[t]{l}CDF\end{tabular}}}}}%
    \put(0,0){\includegraphics[width=\unitlength,page=12]{delays_paper.pdf}}%
    \put(0.24587339,0.50871118){\color[rgb]{0,0,0}\makebox(0,0)[lt]{\lineheight{1.25}\smash{\begin{tabular}[t]{l}\scriptsize TCP\end{tabular}}}}%
    \put(0,0){\includegraphics[width=\unitlength,page=13]{delays_paper.pdf}}%
    \put(0.24587339,0.4654115){\color[rgb]{0,0,0}\makebox(0,0)[lt]{\lineheight{1.25}\smash{\begin{tabular}[t]{l}\scriptsize QUIC\end{tabular}}}}%
  \end{picture}%
\endgroup%